\documentclass[12pt]{article}

\usepackage{graphicx} 
\usepackage{amsmath}
\usepackage{subcaption} 
\RequirePackage{xcolor} 
\RequirePackage{amssymb}
\RequirePackage{amsmath} 
\usepackage[macce]{inputenc} 
\usepackage{siunitx} 
\RequirePackage[T1]{fontenc}
\RequirePackage{times}
\RequirePackage{mathptmx}
\usepackage[bookmarks,raiselinks,pageanchor,hyperindex,colorlinks,citecolor=black,linkcolor=black,urlcolor=blue,filecolor=black,menucolor=black]{hyperref}

\makeatletter \@addtoreset{equation}{section}
\makeatother 
\renewcommand{\thefootnote}{\alph{footnote}}

\newcommand{\email}[1]{{\tt\href{mailto:#1}{#1}}} 

\begin{document}
\thispagestyle{empty}
%
 \mbox{} \hspace{1.0cm}
       December 5, 2017
       \mbox{} \hfill BI-TP 2016/07 \hspace{1.0cm}\\
       \mbox{} \hfill \href{http://www.actaphys.uj.edu.pl/fulltext?series=Sup&vol=10&page=669}{\tt{DOI:10.5506/APhysPolBSupp.10.669}} \hspace{1.0cm}\\
       \mbox{} \hfill \href{http://arxiv.org/abs/1608.04239}{\tt{arXiv:1608.04239v1[hep-ph]}}\hspace{1.0cm}\\
\begin{center}
\vspace*{2cm}
\renewcommand\thefootnote{*}
{{\Large\bf Motion of Confined Particles \footnote{Presented at the \emph{10th International Workshop on Critical Point and Onset of Deconfinement - CPOD 2016}, May 30th - June 4th, 2016,
Wroc{\l}aw, Poland.\\ \\ In Honor of Helmut Satz on the occasion of his Eightieth Birthday and the recent awarding of an Honorary Doctor title at the Uniwersytet Wroc{\l}awski.\\ \\
Published in \emph{Acta Physica Polonica B Proceedings Supplement}, \href{http://www.actaphys.uj.edu.pl/findarticle?series=Sup&vol=10&page=669}{Vol. 10 Number 3 (2017) 669-673.}\\} \\}}
\addtocounter{footnote}{-1}
\renewcommand\thefootnote{\alph{footnote}}
\vspace*{1.0cm}
{\large David E. Miller$^{1,2,}$\footnote{\email{dmiller@physik.uni-bielefeld.de} and \email{om0@psu.edu}} and Dirk Rollmann$^{1,}$\footnote{\email{rollmann@physik.uni-bielefeld.de}}
\\}
\vspace*{1.0cm}
${}^1$ {Fakultät für Physik, Universität Bielefeld, D 33501 Bielefeld}\\
${}^2$ {Department of Physics, Pennsylvania State University,
Hazleton Campus, \\
Hazleton, PA 18202, USA \\}
\end{center}
\vspace*{2cm} {\large \bf Abstract \\} 
We carry out numerical evaluations of the motion of classical particles in Minkowski Space $\mathbb{M}^{4}$ which are confined to the inside of a bag.
In particular, we analyze the structure of the paths evolving from the breaking of the dilatation symmetry, the conformal symmetry and the combination of both together.
The confining forces arise directly from the corresponding nonconserved currents. We demonstrate in our evaluations that these particles under certain initial conditions
move toward the interior of the bag.
\setcounter{footnote}{0}
\renewcommand\thefootnote{\arabic{footnote}}
\newpage
\section{Introduction}
The problem of the confinement of quarks and gluons has been discussed at many different levels. Here we investigate the motion of a particle which is confined inside a region from the inward pressure $B$ of the surrounding vacuum with a fixed energy density $B$.
These properties are basic to the Bag Model \cite{chodos}. In the following we limit our discussion to the consequences of this structure
on the motion of a single particle (parton) confined to the interior of the bag.
Next we discuss the dilatation current $D^\mu(x)$ and the conformal currents $K^{\mu \alpha}(x)$ for $\mathbb{M}^4$. Due to the confining conditions in the Bag Model
the presence of a finite trace for the energy-momentum tensor leads to the breaking of the dilatation and conformal symmetries.
In the following sections we show how this fact changes the motion.
\section{Fundamental Currents}
In this section we describe the dilatation current $D^\mu(x)$ and the four conformal currents  $K^{\mu \alpha}(x)$ in relation of the energy-momentum
tensor $T^{\mu\nu}(x)$ where $\alpha, \mu, \nu = 0,1,2,3$ in terms of the space-time vector $x^\mu \in \mathbb{M}^4$ with a positive-time metric $g_{\mu \nu}$.
The dilatation current is defined as \cite{jackiw}
\begin{equation}
D^{\mu}(x)=x_{\nu}T^{\mu\nu}(x).
\end{equation}
The four conformal currents are written as
\begin{equation}
K^{\mu\alpha}(x)=(2x^{\alpha}x_{\nu}-g^{\alpha}_{\nu}x^2)T^{\mu\nu}(x).
\end{equation}
The basic equations for these currents with finite trace $T^\mu_\mu$ is given in terms of the divergence in $\mathbb{M}^{4}$ as
\begin{eqnarray}
\label{divergenz1}
\partial_{\mu}D^{\mu}(x)=T^{\mu}_{\mu},\\
\label{divergenz2}
\partial_{\mu}K^{\mu\alpha}(x)=2x^{\alpha}T^{\mu}_{\mu}.
\end{eqnarray}
\section{Broken Symmetries}
The presence of a finite trace $T^{\mu}_{\mu} \ne 0$ in Eq. \eqref{divergenz1} and \eqref{divergenz2} provides for broken dilatation and conformal symmetries.
This situation arises from the contrapositive statement of Noether\`{}s Theorem, which yields from the nonconserved currents the broken symmetries.
Thereby we conclude that
\begin{eqnarray}
\partial_{\mu}D^{\mu}(x) \ne 0 \Rightarrow \text{Scalesymmetry is violated,}\\
\partial_{\mu}K^{\mu\alpha}(x) \ne 0 \Rightarrow \text{The conformal symmetry is broken.}
\end{eqnarray}
Physically it is known that for a violation of translational symmetry that forces arose to change the motion.
Similarly for a broken rotational symmetry that a torque was present.
We have previously discussed the implications of Eq. \eqref{divergenz1} and \eqref{divergenz2} for the 1+1 dimensional space-time $\mathbb{M}^{2}$ \cite{rollmann,rollmannMiller}.
\section{Confining Forces inside the Bag}
We recall that the inward bag pressure $B$ and the energy density $B$ gives rise to a diagonal energy-momentum tensor $(+B,-B,-B,-B)$, which
yield
\begin{equation}
\label{spur}
T^{\mu}_{\mu}=g_{\mu\nu}T^{\mu\nu}=4B.
\end{equation}
Next we describe the attractive force along the world line of the moving particle by the \emph{Dyxle Force} $\mathcal{D}(x)$:
\begin{equation}
\mathcal{D}(x)=D^{\mu}x_{\mu}=Bx^{\mu}x_{\mu}=B\tau^2,
\end{equation}
whereby $\tau$ is the proper time. Similarly the \emph{Fourspan} $\mathcal{K}^{\mu\alpha}(x)$ is the combination of the four conformal currents $K^{\mu\alpha}(x)$
so that \cite{miller}
\begin{eqnarray}
\mathcal{K}^{00}(x)=B(2((x^{0})^2-\tau^2));\\
\mathcal{K}^{ii}(x)=B(2(x^{i})^2+\tau^{2})~~\text{with}~~ i=1,2,3;\\
\mathcal{K}^{\iota\kappa}=2Bx^{\iota}x^{\kappa}~~\text{with}~~\iota, \kappa = 0,1,2,3~~\text{and}~~\iota \neq \kappa.
\end{eqnarray}
\section{The Confinement of the Particles inside the Bag}
\begin{figure}[] 
	\centering						
	\includegraphics[width=7cm]{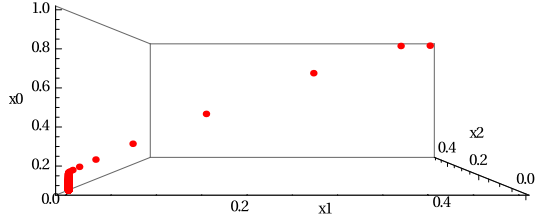}  
	\caption{\label{fig:01}(color online). Three-dimensional representation of the 
trajectory due to only the dyxle force with time axis $x_{0}$, the two spatial 
axes $x_{1}$ and $x_{2}$,  with B=0.1~$\si{\GeV^4}$. The initial values are $x_{0}=1.0$;~$x_{1}=x_{2}=x_{3}=0.5$, in fm.}
\end{figure}
For the formation of the Hadrons in the Bag Model \cite{chodos, satz} we take $M_{h}$ as the hadronic mass and $V_{h}$ as the
hadronic volume. Thus the hadronic energy density $\epsilon_{h}$ is given by
\begin{equation}
\epsilon_{h} = M_{h}/V_{h}.
\end{equation}
In the case of the Bag Model it has been pointed out \cite{chodos,satz} that
\begin{equation}
\label{4B}
\epsilon_{h}= 4B.
\end{equation}
It has been stated above that the hadronic energy density inside the bag is $B$. In addition it has been shown that there are three parts
of the parton kinetic energy density adding up to $3B$. Thus putting it together for the Bag Model, from Eq. \eqref{spur} and \eqref{4B}, we
find that
\begin{equation}
T^{\mu}_{\mu} = 4B = \epsilon_{h}.
\end{equation}
\section{Numerical Evaluation of the Motion in $\mathbb{M}^4$}
\begin{figure}[] 
	\centering
	\begin{subfigure}[]{6cm} 
	\includegraphics[width=6cm]{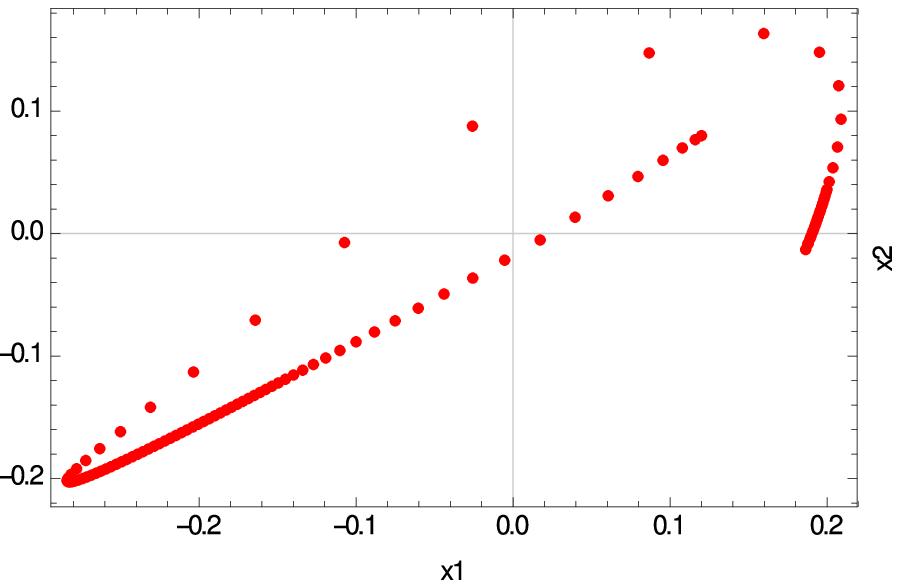} 
	\end{subfigure}
	\quad
	\begin{subfigure}[]{6cm} 
	\includegraphics[width=6cm]{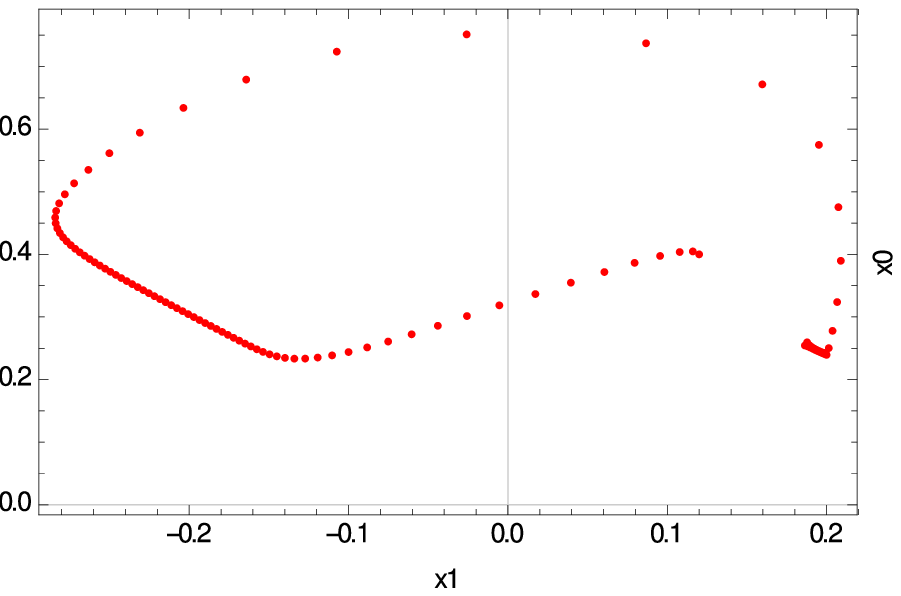}
	\end{subfigure}
	\caption{\label{fig:02}(color online). Two-dimensional representations 
of the trajectory due to only the fourspan with the spatial
axes $x_{1}$ and $x_{2}$ (left) and with the space-time axes $x_{1}$ and $x_{0}$ (right), with B=0.01~$\si{\GeV^4}$. The initial values are 
$x_{0}=0.4$, $x_{1}=0.12$, $x_{2}=0.08$, $x_{3}=0.06$, in fm.}
\end{figure}
\begin{figure}[b] 
	\centering
	\begin{subfigure}[]{5.6cm} 
	\includegraphics[width=5.6cm]{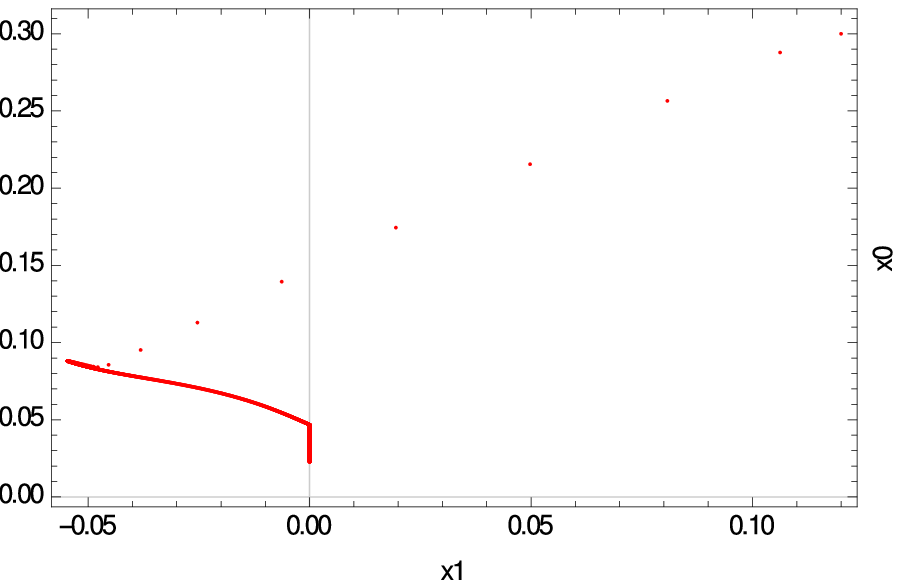} 
	\label{fig:dyxle}
	\end{subfigure}
	\quad
	\begin{subfigure}[]{5.8cm} 
	\includegraphics[width=5.8cm]{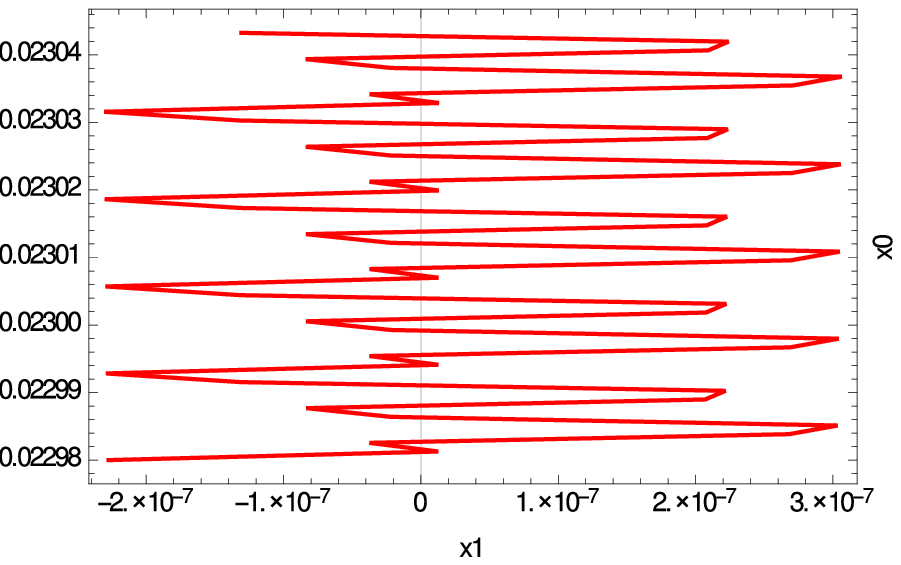}
	\label{subfig:twospan}
	\end{subfigure}
	\caption{\label{fig:03}(color online). Two-dimensional representation 
of the trajectory due to the dyxle and the fourspan together 
with the time axis $x_{0}$ and the spatial axis $x_{1}$, with B=0.1 
$\si{\GeV^4}$. Full range of date (left) and the last 50 values (right). The initial values are 
$x_{0}=0.3$, $x_{1}=0.12$, $x_{2}=0.08$, $x_{3}=0.1$, in fm.}
\end{figure} 
We calculated the trajectories of the discussed forces in a semi-classical fashion. 
The trajectory of the dyxle force is shown in Fig. \ref{fig:01}.
The data show, that the parton moves from the outer light-cone in the interior of the bag.
The trajectory is first more directed to the time axis as an effect of the Minkowski metric.
As a comparison in $\mathbb{M}^{2}$ we can compare in \cite{rollmannMiller} the Fig. 1 (left), for which our
present results show a strong similarity in the plane between the $x_{1}$ and $x_{2}$ axes.
In both the cases of $\mathbb{M}^{2}$ and $\mathbb{M}^{4}$
the dyxle force converges rapidly toward the origin of the confined system.\\
\hspace*{0.5cm}The effect of the fourspan is shown in Fig. \ref{fig:02}.
The two-dimensional plot shows, that the parton moves between 
positive and negative values of the spatial axes $x_{1}$ and $x_{2}$ (left). 
From the calculated data one can see, that the parton moves aside 
and is forced on a kind of rounded orbit in four-dimensional space-time. 
The dependence of the fourspan on the time also causes a loop in 
the time direction (right). The comparison of Fig. \ref{fig:01} (right) in \cite{rollmannMiller} 
brings out clearly the additional structure of $\mathbb{M}^{4}$ of the fourspan.\\
\hspace*{0.5cm}The physically relevant situation is the 
combination of dyxle force and fourspan, since the trace anomaly breaks 
both symmetries simultaneously, see Fig \ref{fig:03}.
The trajectory makes only one movement to the negative spatial 
axis and then the overwhelming dyxle force pulls the parton toward the time 
axis and further toward the origin (left). The fourspan component of these combined forces posses
a permanent fluctuation around the time axis (right).
\section{Summary and Conclusions}
In this work we have presented our results on the motion of the internal constituents (partons) inside the confining structure of the bag.
Whereby we have seen how the presence of the Bag pressure has led to symmetry breakings which brought the presence of
physical forces that altered the parton motion. Under certain conditions we have found that the resulting motion is toward
the properly chosen center of the bag.\\ \\
We would like to thank the organizers for the invitation to CPOD 2016 in Wroc{\l}aw.

\end{document}